\documentclass[12pt]{iopart}
\usepackage{amssymb}
\usepackage[pdftex]{graphicx}
\usepackage{bm}
\begin{document}

\title{ac susceptibility and $^{51}$V NMR study of MnV$_2$O$_4$}

\author{S.-H. Baek$^{1,2}$, K.-Y Choi$^1$, A.P. Reyes$^1$, P.L. Kuhns$^1$,
N.J. Curro$^2$, V. Ramanchandran$^1$, N.S. Dalal$^1$, H.D. Zhou$^1$,
and C.R. Wiebe$^1$}
\address{$^1$National High Magnetic Field Laboratory, Tallahassee, Florida,
32310}
\address{$^2$Los Alamos National Laboratory, Los Alamos, NM, 87545}

\date{\today}

\begin{abstract}
We report $^{51}$V zero-field NMR of manganese vanadate spinel of
MnV$_2$O$_4$, together with both ac and dc magnetization
measurements. The field and temperature dependence of ac
susceptibilities show a reentrant-spin-glass-like behavior below the
ferrimagnetic(FEM) ordering temperature. The zero-field NMR spectrum
consists of multiple lines ranging from 240 MHz to 320 MHz. Its
temperature dependence reveals that the ground state is given by the
simultaneous formation of a long-range FEM order and a short-range
order component. We attribute the spin-glass-like anomalies to
freezing and fluctuations of the short-range ordered state caused by
the competition between spin and orbital ordering of the V site.
\end{abstract}


\submitto{\JPCM}

\section{Introduction}

For several decades, transition-metal spinels of AB$_2$X$_4$ type
have been a subject of active research because of the intriguing
nature of the underlying physics \cite{radaelli05}. The many unusual
phenomena found in the spinel compounds are due primarily to
the corner-shared tetrahedral network of the B cations which resembles
the geometrically frustrated pyrochlore lattice. The magnetic
properties of spinels rely heavily on whether the B cations have
orbital degrees of freedom, and  whether the A cations are magnetic
or nonmagnetic.

Among the vast number of spinels, manganese vanadate MnV$_2$O$_4$
\cite{plumier89,adachi05} is one of the fascinating but very complex
compounds to understand.   This is because (i) the V$^{3+}$ ion
($3d^2$, $S=1$) possesses orbital degeneracy in its $t_{2g}$
orbital; the orbital ordering being a common feature found in other
insulating vanadates (A = Zn \cite{lee04}, Mg \cite{mamiya97}, Cd
\cite{onoda03}); and (ii) both the A and B sites are magnetic as in
manganese chromite, MnCr$_2$O$_4$ \cite{hastings62,tomiyasu04}.

One may expect that MnV$_2$O$_4$ shares common features with
MnCr$_2$O$_4$ as well as with other vanadium spinels. Indeed,
MnV$_2$O$_4$ undergoes long-range FEM ordering at 56 K and
the spin configuration of the B sites changes from collinear to
non-collinear triangular type at low temperature ($T$), as found in
MnCr$_2$O$_4$ \cite{tomiyasu04}. Like other vanadate spinels,
MnV$_2$O$_4$ also undergoes two successive phase transitions --
orbital order and magnetic long-range order. Interestingly, in the
case of MnV$_2$O$_4$, the cubic-to-tetragonal structural transition
related to the orbital order takes place at $T_S=53$ K, just below
the FEM transition temperature $T_C=56$ K \cite{adachi05}.
Here the question arises about the combined effect of the two
features (i) and (ii) on a ground state. However, this issue has not
yet been thoroughly addressed.

In this paper, we report the observation of the complex ground state
in MnV$_2$O$_4$ through ac/dc magnetization and  zero-field (ZF)
$^{51}$V NMR measurements.  Our data reveal that there exists a
short-range ordered state on top of the long range FEM order,
leading to a spin-glass-like behavior. They are discussed in terms
of the competing spin and orbital ordering of the V$^{3+}$ ion.

\section{Sample preparation and experimental details}

Polycrystalline powder sample of MnV$_2$O$_4$ was prepared by
standard solid-state reaction. Stoichiometric mixtures of MnO and
V$_2$O$_3$ were ground together and pressed into pellet, the pellet
was placed into an evacuated quartz tube ($\sim 10^{-5}$ torr) and
fired for 40 hours at 950 $^\circ$C. Samples were characterized by
X-ray powder diffraction. Both dc and ac  susceptibility
measurements were performed as a function of both field and
temperature using SQUID magnetometer (Quantum Design MPMS). An
applied frequency was in the range of $10^{-1}-10^{3}$ Hz with a
driving field magnitude of 5 Oe. NMR experiments were carried out at
zero external field using a coherent pulsed spectrometer capable of
computer-controlled tuning of a probe which is calibrated over a
wide frequency range. In preliminary measurements using an untuned
NMR probe ($Q=1$), we detected two groups of signals at near 280 MHz
and 560 MHz at 4.2 K. These were assigned to the NMR signals of
V$^{3+}$ and Mn$^{2+}$ ions, respectively. A probe with tank circuit
was subsequently employed to focus on the stronger signal of the V
site. The $^{51}$V NMR spectra were obtained by integrating averaged
spin echo signals as the frequency was swept through the resonance
line.

\section{Results and discussion}
\subsection{Magnetic susceptibilities}

\begin{figure}
\label{fig:1} \centering
\includegraphics[width=3.5in]{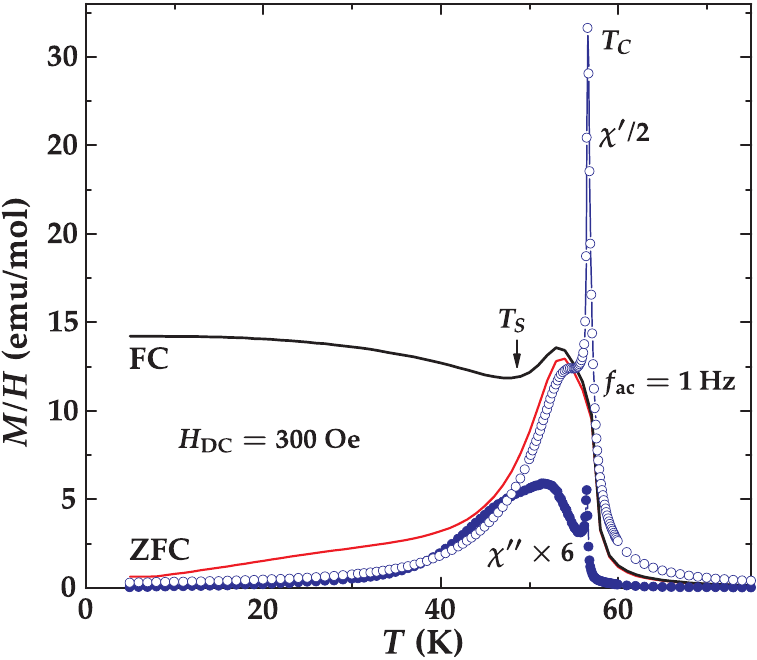}
\caption{(Online color) Temperature dependence of the
ac (circle symbols) and dc (solid lines) susceptibilities of
MnV$_2$O$_4$. For clarity, the real and imaginary part of the ac
susceptibilities are divided by 2 and multiplied by 6,
respectively.}
\end{figure}

Figure 1 shows the temperature dependence of the dc susceptibility
under an applied field of 300 Oe together with the ac susceptibility
at a frequency of 1 Hz. The field-cooled (FC) and zero-field-cooled
(ZFC) curves display a pronounced bifurcation below the FEM ordering
temperature of $T_C$. As $T_C$ is approached from above, the FC
curve increases steeply and then shows a dip around the structural
phase transition temperature at $T_S$  and finally increases
monotonically. In contrast, the ZFC curve exhibits a maximum and
then a monotonic decrease upon cooling from $T_C$. Our results are
in a good agreement with Ref.~\cite{adachi05}. Noticeably, one
observes a substantial difference between the ac and ZFC dc
susceptibilities below $T_C$, implying the presence of
reentrant-spin-glass (RSG)-like anomalies.

\begin{figure}
\label{fig:2} \centering
\includegraphics[width=4in]{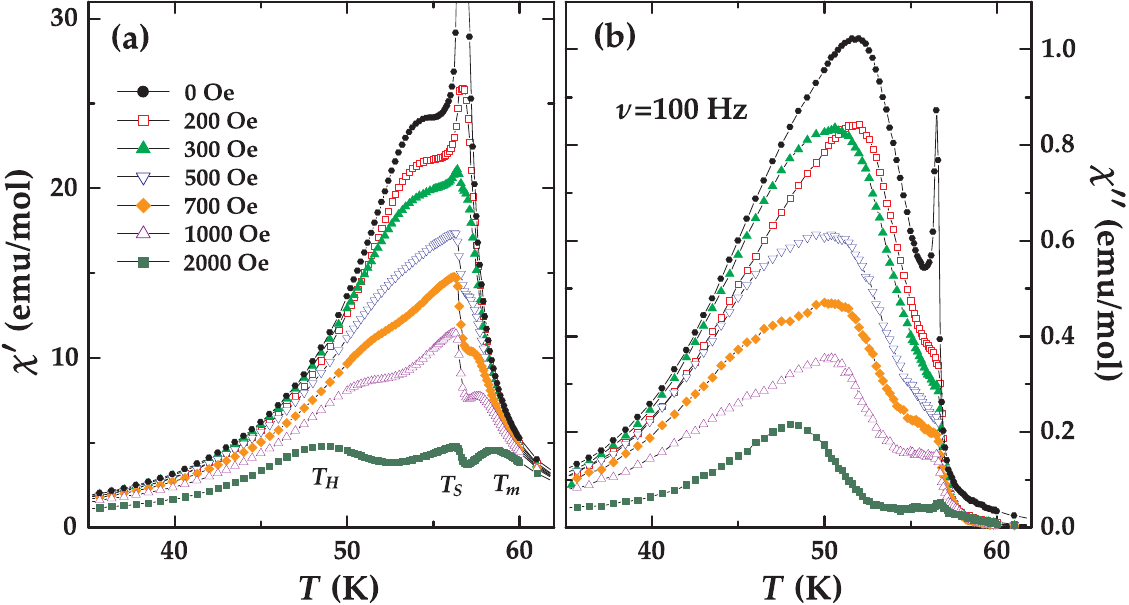}
\caption{(Color online) Field dependence of the real (a) and
        imaginary (b) parts of the ac susceptibility of MnV$_2$O$_4$
as a function of temperature. A field is driven at H$_{ac}=5$ Oe and
$\nu=100$ Hz. }
\end{figure}

In figure 2 we present the temperature dependence of the ac
susceptibilities, $\chi(H_a,T)$ for different static field $H_a$
applied parallel to the ac driving field. With decreasing
temperature the real part of $\chi(H_a,T)$ exhibits a sharp peak
around $T_C$ and then a round maximum and finally falls off upon
further cooling. With increasing $H_a$, the latter maximum
(designated by $T_H$) decreases in both amplitude and temperature.
This so-called Hopkinson maximum is not critical in origin, but
arises from processes associated with the regular/technical
contributions to the susceptibility (for example, domain wall motion
and coherent rotation) \cite{chikazumi97}. The former sharp peak
develops to two peaks (denoted by $T_m$ and $T_S$, respectively)
with increasing dc field. The peak $T_m$ is suppressed in amplitude
and shifts to higher temperature with an increase in the dc field.
This is due to critical fluctuations accompanying a transition from
a paramagnetic to a FEM state \cite{williams91}. In contrast, the
peak $T_S$  decreases in amplitude without showing any shift in
temperature. This might be associated with the structural phase
transition~\cite{adachi05}, leading to a pinning of domain walls.
The imaginary part of $\chi(H_a,T)$ shows a sharp peak around $T_C$,
which exhibits a similar behavior as the peak $T_S$. In addition,
the loss peak of $\chi''(H_a,T)$ shows up around 51 K, which is
suppressed in both temperature and amplitude as $T_a$ increases.

\begin{figure}
\label{fig:3} \centering
\includegraphics[width=4in]{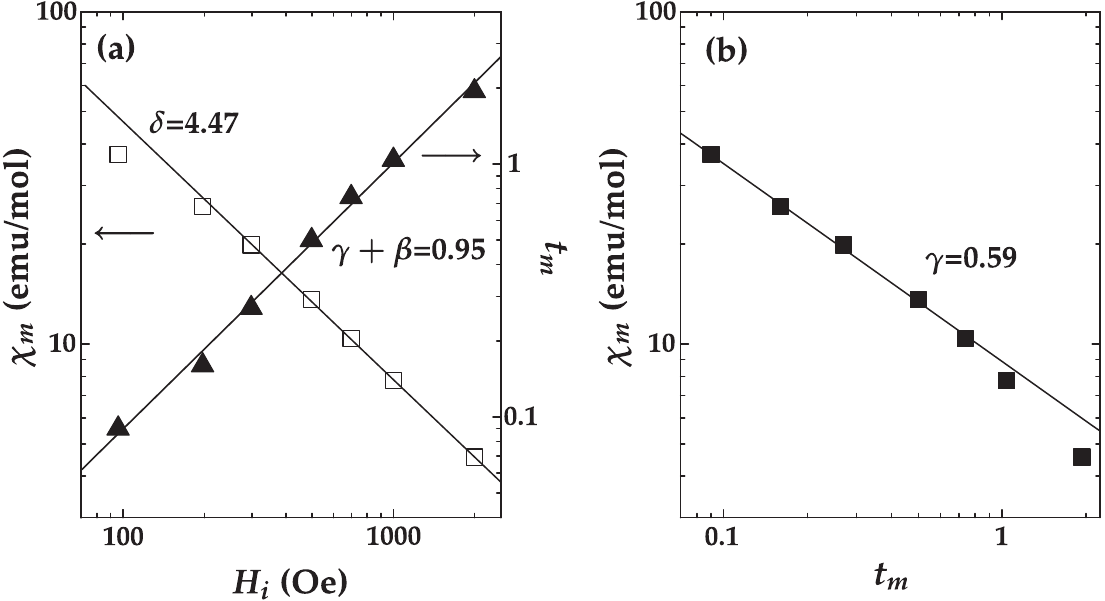}
\caption{(a) The critical field amplitude versus the internal
        field (open square) is presented together with the reduced temperature
        versus the internal field (full triangle)
        on a double-logarithmic scale. (b) A double-logarithmic plot of the
        critical field temperature versus the reduced temperature. The solid
        and dotted lines are a fit to Eqs. (1)-(3).}
\end{figure}

Since the peak $T_m$ is governed by critical fluctuations, it is
expected to follow the static scaling law. The field and temperature
dependence of the critical peak is related to the standard critical
exponents \cite{stanley71,campostrini02,Zhao99} by

\begin{eqnarray}
\chi(H_i,T_m) & \propto  & H_i^{1/\delta -1} \\
t_m & \propto & H_i^{(\gamma + \beta)^{-1}} \\
\chi(H_i,T_m) & \propto & t_m^{-\gamma},
\end{eqnarray}
where $H_i$ is the internal field given by $ H_a=H_i- NM$ where $N$
is the demagnetizing factor and $M$ is the magnetization. In our
case, $N$ is estimated by the slope of the low field shearing curves
around $T_C$ (not shown here). $t_m$ is the reduced temperature
defined as $t_m=(T_m - T_C)/T_C$. We determine the temperature of
$T_C=56.58$ K from the zero field data. The critical field amplitude
$\chi(H_i,T_m)$ is directly obtained from Fig. 2(a). In Fig. 3 we
provide a double-logarithmic plot of $\chi(H_i,T_m)$ versus $H_i$,
$t_m$ versus $H_i$, and $\chi(H_i,T_m)$ versus $t_m$, respectively. A
close look at the data reveals a small curvature at low fields. This
is typical for the system with exchange coupling strength disorder
\cite{williams91}. A least squares fit of the data between the
restricted fields yield the exponent values of $\delta=4.47 \pm
0.05$, $\gamma + \beta=0.95 \pm 0.04$ and $\gamma=0.59 \pm 0.03$.
The $\delta$ and $\beta$ values are close to the prediction of a 3D
Heisenberg model \cite{campostrini02}. However, the $\gamma$ value
strongly deviates from it. Although we cannot exclude  the
uncertainties in evaluating exponents, it might be ascribed to a
first-order-like transition due to strong spin-orbital coupling as
well as to a competition between frustration and a structural phase
transition (see below).

\begin{figure}
\label{fig:4} \centering
\includegraphics[width=4in]{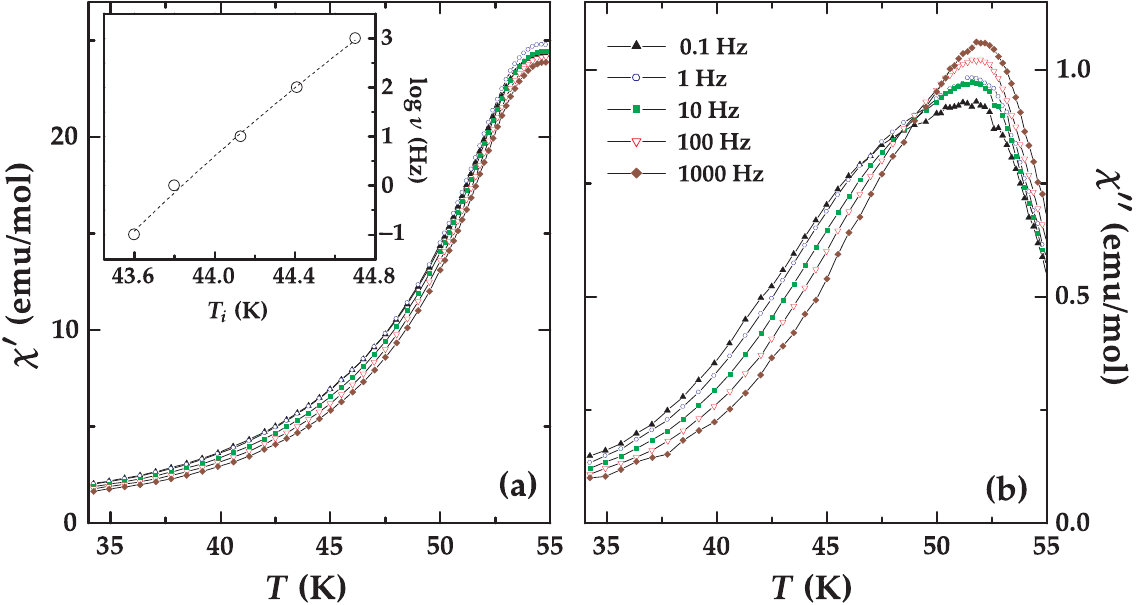}
\caption{(Color online) Frequency dependence of the real (a) and
imaginary (b) parts of the ac susceptibility as a function of
temperature. Inset: Frequency shift of the inflection point of the
real part of the susceptibility versus temperature.}
\end{figure}

In figure 4 the temperature dependence of the $\chi'(T)$ and
$\chi''(T)$ is presented as a function of frequency in the range of
$10^{-1} - 10^3$ Hz. Both  $\chi'(T)$ and $\chi''(T)$ exhibit
pronounced frequency dependence between 20 K and $T_C$. Overall,
with increasing frequency both  $\chi'(T)$ and $\chi''(T)$ shifts to
higher temperature. This is unexpected for a long-range ordered FEM
state, implying that a spin freezing phenomenon takes place in the
FEM ordered background.

The shift of the ac susceptibility as a function of the frequency is
estimated by taking the inflection point of $\chi'(T)$ [see the
inset of figure 4(a)]. From the maximal shift $\Delta T_i$ we obtain
the value of $\varphi=\Delta T/T_i \Delta {\mbox( \log} \omega
{\mbox ) }\sim 0.007$. For a spin glass the value is typically of
the order of $10^{-2}$, while for a superparamagnet it lies between
$10^{-1}-10^{-2}$ \cite{mydosh}. Using this criterion, we conclude
that the studied compound belongs to a class of a spin glass.
Further, we fit the experimental data in terms of the Vogel-Fulcher
law $\omega=\omega_0 \exp[-E_a/k_B(T_i - T_0)]$ to figure out the
nature of the spin freezing process. It yields the values of
$\omega_0 \approx 10^{33}$ Hz, $E_a\approx 287$ K, and $T_0\approx
33.5$ K. The unrealistic large value of $\omega_0$ suggests that the
spin freezing is not of simply cooperative but governed by an
intricate process.

\subsection{NMR measurements}

$^{51}$V zero-field NMR spectrum and its temperature evolution are
shown in figure 5(a) for slow cooling (SC), with the rate of $\sim$1
K/min, and in figure 5(b) for rapid cooling (RC), with the rate of
60 K/min. The Boltzmann correction for the signal intensities  has
been made by multiplying each spectrum by $T$. The spectrum has a
complex structure and spans a wide range of frequencies from 240 to
320 MHz. The resonance frequency in zero field is given as $\gamma_N
H_{\mathrm{hf}}$ where $\gamma_N$ ($=11.193$ MHz/T) is the nuclear
gyromagnetic ratio. The hyperfine field $H_{\mathrm{hf}}$ is dominated
by the core polarization of the inner $s$-electrons by the outer
unpaired $d$-electrons. For the V$^{3+}$ ion ($3d^2$, $S=1$), the
estimated $H_{\mathrm{hf}}$ due to pure Fermi contact is about $\sim
25$ T \cite{freeman65} which falls into $\sim 280$ MHz range. The
contributions from transferred, dipolar, or orbital hyperfine fields
are negligible.

One can infer from figure 5 that the $^{51}$V NMR line is composed
of several overlapping structures, whose peaks are somewhat
resolved. This is not surprising given the low crystal symmetry
\cite{footnote}. One possibility is the quadrupole interactions
split the $^{51}$V ($I=7/2$) degenerate levels producing $2I=7$
transitions. However we discard this model since the quadrupole
coupling required to represent the spectrum would be too large for
the V nucleus.   Hence, we attribute the structures to V sites
belonging to several magnetically differentiated sites.

\begin{figure}
\label{fig:5}
\centering
\includegraphics[width=4in]{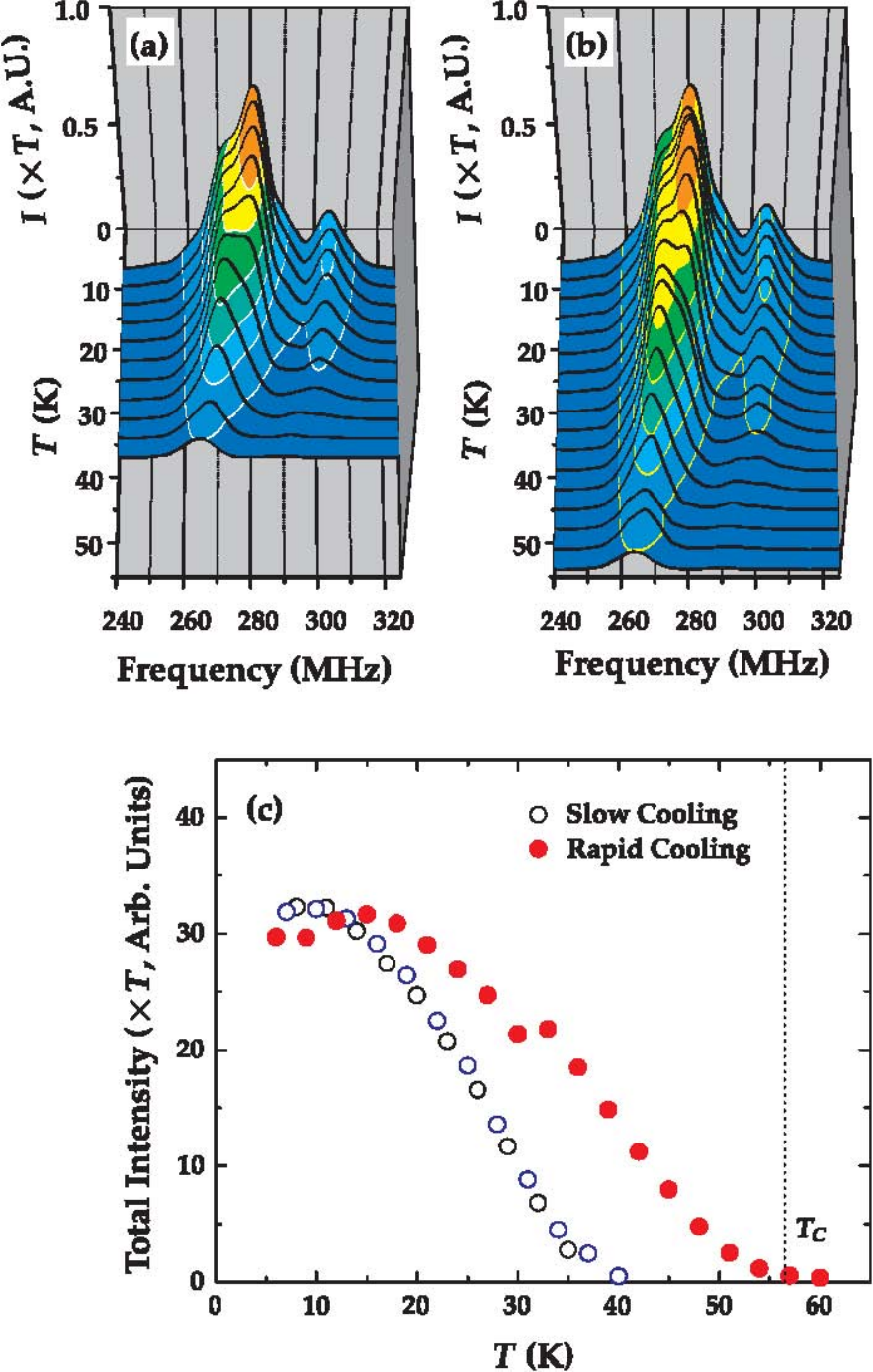}
\caption{(Online color) (a) Evolution of the NMR spectrum with increasing $T$
in ZF for slow cooling condition.
The Boltzmann correction for intensity was taken into account by scaling each
spectrum by $T$. (b) For rapid cooling, the evolution of the spectral shape is
the same except the observation of the signal at much higher temperature. (c)
Total integrated intensity of the NMR signal in zero field was plotted
against temperature for slow and rapid cooling. Note that, for rapid cooling, the signal was
detected even above $T_C$.}
\end{figure}

The temperature dependence of the integrated intensity multiplied by
$T$ of the spectrum is shown in figure 5 (c) for both SC and RC. The
signal for the RC case can be observed even above $T_C$ while the
signal for SC becomes very weak near 40 K. The loss of NMR signal
intensity at high $T$ is due to the fast spin fluctuations i.e.,
shortening of $T_2$ associated with magnetic ordering. The different
temperature dependence of the signal intensity for different cooling
rate points to the effects of disorder which influences the freezing
rate of the short-range correlated spins, a conclusion also deduced
previously from the magnetization results. At low $T<20$ K, however,
the signal intensity and shape are the same regardless of the
cooling rate implying that the spin ground state of this compound is
not affected by the thermal history. This is consistent with the ac
susceptibility results, which show no frequency dependence at the
same temperature range.

By examining the evolution of the spectrum in figure 5 (a) and (b),
we take note that the NMR peak near 265 MHz is quite distinguishable
from the other lines, and survives after the disappearance of the
other lines at high temperature. Also we see that the shift of the
NMR line is very small, if any, up to temperature near $T_C$. This
is surprising since the NMR resonance frequency usually follows the
bulk magnetization which often shows gradual second order transition
behavior. Therefore, the NMR data suggest that the magnetic
transition at $T_C$ must be of first order.

In both SC and RC cases,
the stable configuration is established below  15 K as shown in figure 5.
Here we show that the intriguing ground state is expected in the
presence of several competing interactions. Spinels contain two
different exchange coupling constants: $J_{\mathrm{AB}}$ and
$J_{\mathrm{BB}}$ between the A and B and the B and B sites,
respectively. For a cubic phase, a magnetic configuration relies on
the parameter $u$  defined as \cite{menyuk62}
\begin{equation}
\label{eq1}
        u = \frac{4J_{\mathrm{BB}} S_{\mathrm{B}}}{3J_{\mathrm{AB}}S_{\mathrm{A}}},
\end{equation}
where $S_{\mathrm{A}}$ and $S_{\mathrm{B}}$ are spin magnitudes at the
respective A and B sites. The relative strength of two interactions
determines a ground state. For example, the ground state of
chromites [Co,Mn]Cr$_2$O$_4$ with $u=1.5-2.0$ shows a coexistence of
FEM long-range order and spiral short-range order that
causes the RSG-like behavior \cite{tomiyasu04}. The MnV$_2$O$_4$
system seems to lie in the same parameter range as suggested by the
fact that the RSG-like behavior starts to appear at $T_C$  of the
cubic phase (see figure 1).

In [Co,Mn]Cr$_2$O$_4$ the appearance of the short-range order is
ascribed to the residual magnetic frustration in the B site, which
is non-vanishing when $J_{\mathrm{AB}}$ is smaller than
$J_{\mathrm{BB}}$. In our case, however, this scenario seems to be
less plausible since the frustration could be relived through the
orbital ordering of the V$^{3+}$ ions. Rather, the short-range
character in MnV$_2$O$_4$ may be due to the specific orbital
ordering that occurs at $T_S$. In the cubic phase ($T>T_S$) the
$t_{2g}$ orbitals are degenerate and thus the orbital degrees of
freedom are irrelevant. The V sublattice aligns ferromagnetically
due to the antiferromagnetic couplings between V$^{3+}$ and
Mn$^{2+}$ ions. In contrast, in the tetragonal phase ($T<T_S$), a
structural phase transition takes place to one whose orbital
configuration is intimately coupled to a spin configuration via a
orbital-spin coupling~\cite{adachi05}. Accordingly, an
antiferro-orbital arrangement of the V$^{3+}$ ions is preferred due
to the dominant ferromagnetic interaction between V$^{3+}$ ions.

This staggered-type orbital configuration of MnV$_2$O$_4$ is
contrasted to the orbital patterns found in vanadium spinels
(A=Zn,Mg,Cd), which are the nonmagnetic analogue of MnV$_2$O$_4$ in
the A site \cite{tsunetsugu03,tchernyshyov04, matteo05}. Tsunetsugu
and Motome \cite{tsunetsugu03} suggested alternating
$d_{xy}$--$d_{xz}$ and $d_{xy}$--$d_{yz}$ orbital chain structure,
which is compatible with the space group $I4_1/a$. On the other
hand, Tchernyshyov \cite{tchernyshyov04} argued against the above
model and proposed the space group $I4_1/amd$ which is associated
with a complex linear combination of $d_{xz}$ and $d_{yz}$ orbitals.
Regardless of the detailed picture, the one-dimensional
antiferromagnetic chain is the essential ingredient to both models
owing to the direct overlap of $d_{xy}$ orbitals \cite{lee04}. This
means that the orbital ordering imposed by the exchange couplings of
V$^{3+}$---Mn$^{2+}$ ions is different from the orbital pattern
favored by the structural phase transition. This causes the
instability of the antiferro-orbital configuration. Although
frustration is relieved through the structural phase transition, it
also paves the way for the competing orbital ordering of V$^{3+}$
ions to occur.

\begin{figure}
\label{fig:6}
\centering
\includegraphics[width=4in]{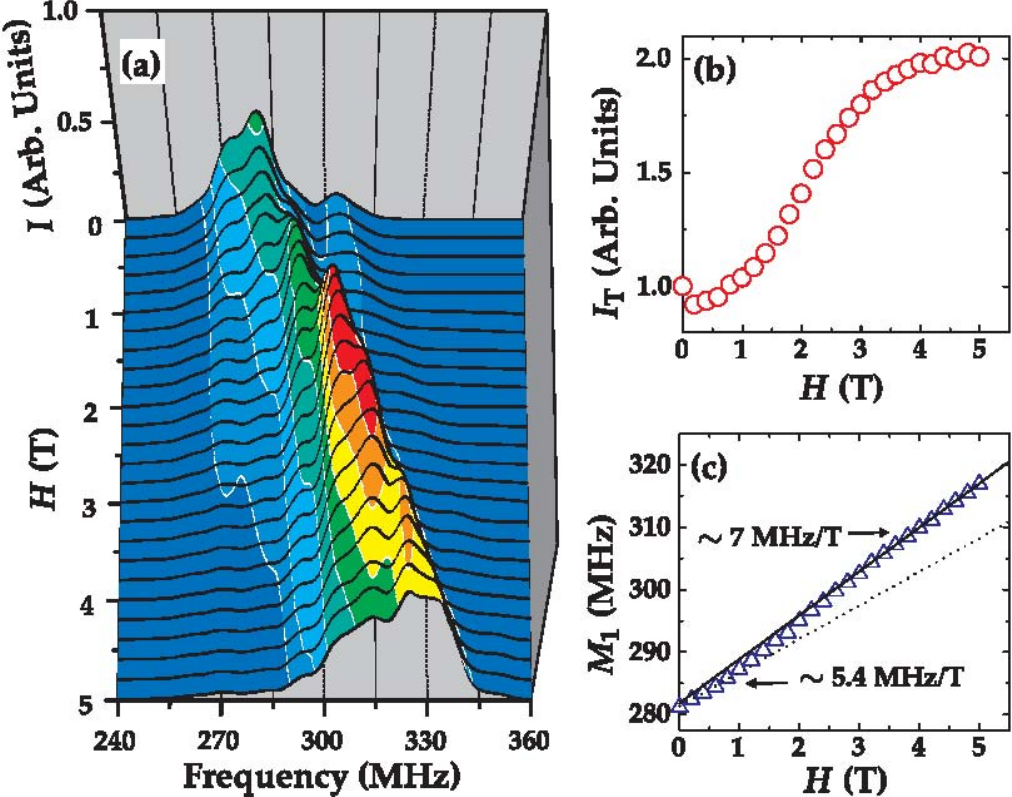}
\caption{(Online color) (a) External field dependence of spectrum up to 5 T.
(b) Total integrated intensity increases with external field saturating at
$\sim$ 4 T. (c) First moment versus
external field. The initial slope of 5.4 MHz/T changes to 7 MHz/T at higher
fields.  The value must be compared to the expected value of 11.193 MHz/T.}
\end{figure}

Lastly, we present the external field ($H$) dependence of the NMR
spectrum in figure 6. We observe that each identified line at zero
field does not shift at all with $H$.  Instead, amplitude of each
peak decreases or increases without shift in such a way that the
first moment ($M_1$) or the center of gravity of the spectrum
changes linearly with increasing $H$ as shown in figure 6 (c).
Moreover, new NMR lines appear as $H$ increases. The slope of $M_1$
versus $H$ is initially 5.4 MHz/T but get bigger smoothly up to 3 T
being fixed to  7 MHz/T. The value is still smaller than the
gyromagnetic ratio $\gamma_N$ of V (11.193 MHz/T). We conjecture that
this behavior may be possible only with assumption of the discrete
spin directions which is distributed spatially. The change of the
slope below 3 T may be due to the effect of the rotation of the
domain or the similar stabilized process against the external field.

Also we observed that the total intensity of the spectrum increases
with increasing $H$ up to 4 T above which it is saturated [see
figure 6 (b)]. Typically, NMR signal enhancement is associated with
the domain walls which are swept out with $H$ resulting in the
disappearance of the enhancement effect. In this regard, the usual
enhancement of the signal with large $H$ suggests the persistence of
the domain motions at least up to 4 T. Since the domains of the
long-range ordered phase are expected to be frozen-in at low fields,
the enhanced high-field signal might be related to the domain
motions of the short-range ordered state. Above we have ascribed the
short-range order to the orbital ordering induced by the structural
phase transition. Specifically, the V ions form the
antiferromagnetic chain with the exchange coupling constant of an
order of 1000 K \cite{adachi05}. In this situation, the short-range
ordered component of the V spins is frustrated. Thus, in the studied
field range the short-range ordered state undergoes a minor change
in the spin directions accompanying the domain motions. To draw the
more concrete picture on the puzzling behavior of the field
dependence, a further study of the single crystal is needed.

Within our study, we cannot determine the exact nature of the
short-range ordered ground state. More direct determination has to
come from a neutron diffraction study.
 However, it cannot be
identified as a classical spin glass phase because its NMR signal
intensity is comparable to that of the long-range ordered phase. It
is possible that the RSG-like behavior is associated with the FEM
domains and is a consequence of the freezing of fluctuations of the
short-range ordered state.

\section{Conclusion}

In conclusion, we examined the ground state of MnV$_2$O$_4$ through
magnetization and NMR measurements. We observed the reentrant-spin-glass-like behavior
on top of the long-range FEM order.
Our study provides a clear-cut example of the inhomogeneous ground
state in {\it undoped} compounds where spin interactions and orbital
ordering related to a structural phase transition are two competing
parameters.

\ack This work was supported by NSF in-house research program State
of Florida under cooperative agreement DMR-0084173.

\section*{References}
\bibliography{mnv2o4_nmr_ac}

\end{document}